\documentclass[11pt]{article}
\usepackage[final, nonatbib]{neurips_2023} 
\usepackage[dvipsnames]{xcolor}

\usepackage{bbm}
\usepackage{amssymb}
\usepackage{amsmath}
\usepackage{amsthm}
\usepackage{aliascnt}
\usepackage{mathtools}
\usepackage{biblatex}
\usepackage{svg}
\usepackage{graphicx}
\usepackage{subcaption}
\usepackage{mathrsfs}
\usepackage[bb=dsserif]{mathalpha}
\usepackage{xr-hyper}

\usepackage{hyperref}

\theoremstyle{plain}
\newtheorem{theorem}{Theorem}[section]

\newaliascnt{proposition}{theorem}

\aliascntresetthe{proposition}

\newaliascnt{corollary}{theorem}

\aliascntresetthe{corollary}

\newaliascnt{lemma}{theorem}
\newtheorem{lemma}[lemma]{Lemma}
\aliascntresetthe{lemma}

\theoremstyle{definition}
\newtheorem{definition}{Definition}[section]

\theoremstyle{remark}
\newtheorem{rmk}{Remark}[section]

\newtheorem*{rmk*}{Remark}
\newtheorem*{fact*}{Fact}
\newtheorem{state}{Statement}[section]

\newenvironment{remark}
{\pushQED{\qed}\rmk}
{\popQED\endrmk}

\input{preamble.sty}
\title{Model-adapted Fourier sampling for generative compressed sensing}

\author{%
  Aaron Berk\thanks{Dept.\ Mathematics and Statistics, McGill University, Montr\'eal, Canada. Now affiliated with Deep Render, London UK}
  \And
  Simone Brugiapaglia\thanks{Dept.\ Mathematics and Statistics, Concordia University, Montr\'eal, Canada}
  \And
  Yaniv Plan \And Matthew Scott\thanks{Corresponding author: \texttt{matthewscott@math.ubc.ca}}  $\qquad$ Xia Sheng \thanks{Authors listed in alphabetic order.  MS was primarily responsible for developing the theory and writing the main body, XS was primarily responsible for numerical experiments and writing the numerical section.} \And \"{O}zg\"{u}r Y\i{}lmaz\thanks{YP, MS, XS and OY affiliated with Dept.\ Mathematics, University of British Columbia, Vancouver, Canada}
}

\begin{document}
\maketitle

\begin{abstract}
	We study generative compressed sensing when the measurement matrix is randomly subsampled from a unitary matrix (with the DFT as an important special case). It was recently shown that $\textit{O}(kdn\| \boldsymbol{\alpha}\|_{\infty}^{2})$ uniformly random Fourier measurements are sufficient to recover signals in the range of a neural network $G:\field^k \to \field^n$ of depth $d$, where each component of the so-called local coherence vector $\boldsymbol{\alpha}$ quantifies the alignment of a corresponding Fourier vector with the range of $G$. We construct a model-adapted sampling strategy with an improved sample complexity of $\textit{O}(kd\| \boldsymbol{\alpha}\|_{2}^{2})$ measurements. This is enabled by: (1) new theoretical recovery guarantees that we develop for nonuniformly random sampling distributions and then (2) optimizing the sampling distribution to minimize the number of measurements needed for these guarantees. This development offers a sample complexity applicable to natural signal classes, which are often almost maximally coherent with low Fourier frequencies. Finally, we consider a surrogate sampling scheme, and validate its performance in recovery experiments using the CelebA dataset.
\end{abstract}

\section{Introduction}
\label{introduction}
Compressed sensing considers signals
$\boldsymbol{x}_0 \in \field^n$ with high ambient dimension $n$ that belong to (or can be well-approximated by elements of) a prior set $\mathcal{V} \subseteq \field^n$ with lower ``complexity" than the ambient space. The aim is to recover (an approximation to) such signals with provable accuracy guarantees from the linear, typically noisy, measurements $\boldsymbol{b} = A\boldsymbol{x}_0 +\boldsymbol{\eta}$, where $\boldsymbol\eta \in \measfield^m$ denotes noise and $A \in \measfield^{m \times n}$ with \( m \ll n \) is an appropriately chosen (possibly random) measurement matrix. The signal is to be recovered by means of a computationally feasible method that utilizes the structure of $\mathcal{V}$ and has access to only $A$ and $\boldsymbol{b}$.
In classical compressed sensing, $\mathcal{V}$ is the set of sparse vectors. In generative compressed sensing, the prior set $\mathcal{V}$ is chosen to be the range of a generative neural network $G:\field^k \to \field^n$, an idea that was first explored in~\cite{boraCompressedSensingUsing2017}. 
Our results will hold for ReLU-activated neural networks as defined in~\cite[Definition~I.2]{berkCoherenceParameterCharacterizing2022}. Here we denote the ReLU (Rectified Linear Unit) activation as $\sigma(z):=\max\{z,0\}$, applied component-wise to a real vector $\boldsymbol{z}$.

\begin{definition}[$(k,d,n)$-Generative Network]
	\label{kdn-generative_network}
	With \( k, d, n \in  \mathbb{N} \), fix the integers $2 \leq k := k_{0} \leq k_{1}, \ldots, k_{d-1} \le k_d = n$, and for $i \in [d]$, let
	$W^{(i)} \in \field^{k_{i}\times k_{i-1}}$. A $(k,d,n)$-generative network is
	a function $G : \field^{k} \to \field^{n}$ of the form
	$G(\boldsymbol{z}) := W^{(d)} \sigma \left( \cdots W^{(2)} \sigma \left( W^{(1)} \boldsymbol{z} \right) \right).$
\end{definition}
In the same work~\cite{boraCompressedSensingUsing2017}, the authors provided a theoretical framework for generative compressed sensing
with $A \in \mathbb{R}^{m \times n}$ having independent identically distributed (i.i.d.)\ Gaussian
entries. However, the assumption of Gaussian measurements is unrealistic for applications like MRI, where measurements are spatial Fourier transforms. Limitations of the hardware in such an application restrict the set of possible measurements to the rows of a fixed unitary matrix (or approximately so, up to discretization of the measurements).
Hence we consider the more realistic \emph{subsampled unitary} measurement matrices, i.e., matrices with rows randomly subsampled from the rows of a unitary matrix $F \in \measfield^{n \times n}$.

A first result in the generative setting with subsampled unitary measurements~\cite{berkCoherenceParameterCharacterizing2022} shows that for well-behaved networks,
$m = \bigO(k^2d^2)$ (up to log factors) measurements is sufficient for recovery with high probability.
However, sampling uniformly is not efficient; sampling more informative measurements at a higher rate is
known to improve performance in compressed sensing (e.g., low Fourier frequencies have strong correlation with natural
images, they will therefore tend to be more informative)~\cite{adcockCompressiveImagingStructure2021}. This idea is mirrored in the radial sampling strategy used in MRI scans, which takes a disproportionate number of low-frequency measurements~\cite{muradRadialUndersamplingBasedInterpolation2021}. We address this limitation by generalizing the theory to any measurement matrix of the form $A = SF$ where $F \in \measfield^{n \times n}$ is a unitary matrix and $S \in \mathbb{R}^{m \times n}$ is a \emph{sampling matrix}, which we now define. We adopt the convention that $\boldsymbol{e}_j$ denotes a canonical basis vector and that $\Delta^{n-1}$ is the simplex in $\mathbb{R}^n$. See \autoref{sec:notation} for the definition of symbols throughout this paper.
\begin{definition}[Sampling Matrix]
	\label{sampling_matrix}
	We define a \emph{sampling matrix} to be any matrix $S \in \mathbb{R}^{ m\times n  }$ composed of i.i.d.\ row vectors $\boldsymbol{s}^*_1, \dots, \boldsymbol{s}^*_m$ such that $\mathbb{P}(\boldsymbol{s}_i = \boldsymbol{e}_j)= p_j$ for all $i \in [m], j \in [n]$, for some fixed probability vector $\boldsymbol{p} \in \Delta^{n-1}$.
\end{definition}
We will further show, similarly to~\cite{puyVariableDensityCompressive2011}, that picking the sampling
probabilities in a manner informed by the geometry of the problem yields
improved recovery guarantees relative to the uniform case. Specifically, we provide a bound on the measurement complexity of $m=\bigO(k^2d^2)$ even for models which are highly aligned with a small subset of the rows of $F$. To find good sampling probabilities, we must understand that measurements (i.e. the rows of the measurement matrix) are effective for a prior set $\mathcal{V} \subseteq \field^n$ if they help differentiate between signals in $\mathcal{V}$. Therefore, we consider the alignment of rows of $F$ with the set $\mathcal{V}-\mathcal{V}$ (where the difference is in the sense of a Minkowski sum, see \autoref{sec:notation}). We will sample rows of $F$ that have a high degree of alignment with $\mathcal{V}-\mathcal{V}$ at a higher rate. For technical reasons, we consider alignments with a slightly larger set, given by the following set operator previously introduced in~\cite[Definition~2.1]{berkCoherenceParameterCharacterizing2022}.

\begin{definition}[Piecewise Linear Expansion]
	\label{piecewise_linear_expansion}
	Let $\mathcal{C} \subseteq \field^{n}$ be the union of $N$ convex cones: $\mathcal{C} = \bigcup_{i=1}^N \mathcal{C}_i$. Define the piecewise linear expansion
	$
		\Delta(\mathcal{C}) := \bigcup_{i=1}^N \Span(\mathcal{C}_i) = \bigcup_{i=1}^N
		(\mathcal{C}_i - \mathcal{C}_i).
	$
\end{definition}

The piecewise linear expansion is a well-defined set operator as shown in~\cite[Remark A.2]{berkCoherenceParameterCharacterizing2022}. Specifically, it is independent of the choice of convex cones $\mathcal{C}_i$.

We require the following quantity to quantify the alignment of the individual vectors with the prior set.

\begin{definition}[local coherence]
	\label{local_coherence}
	The \emph{local coherence} of a vector $\boldsymbol\phi \in \measfield^n$ with respect to a cone $\mathcal{T} \subseteq \field^n$  is defined as
	$$\alpha_{\mathcal{T}}(\boldsymbol\phi) := \sup_{\boldsymbol{x} \in \Delta(\mathcal{T}) \cap \sphere{n}} \lvert \langle \boldsymbol\phi, \boldsymbol{x}\rangle \rvert.$$
	The \emph{local coherences} of a unitary matrix $F \in \measfield^{n \times n}$ with respect to a cone $\mathcal{T} \subseteq \field^n$ are collected in the vector $\boldsymbol{\alpha}$ with entries $\alpha_j := \alpha_\mathcal{T}(\boldsymbol{f}_j)$, where $\boldsymbol{f}^*_j$ is the $j^{th}$ row of $F$.
\end{definition}
By using the local coherences of $F$ to inform sampling probabilities, we will show that recovery occurs from only $m = \bigO(kd\lVert \boldsymbol{\alpha}\rVert_2^2)$ (up to log factors) measurements with high probability.

\begin{figure}[h!]
\vspace{-11pt}
	\centering
	\includegraphics[width=1\textwidth]{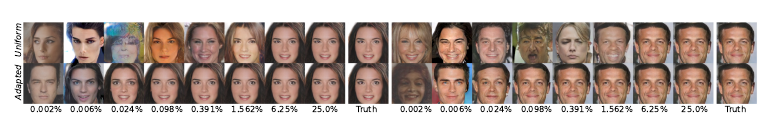}
 
\vspace{-12pt}
	\caption{Samples recovered by different sampling schemes with different sampling rates}
	\label{fig:intro_comp}
\vspace{-12pt}
\end{figure}

\paragraph{Prior Work}
\label{prior_work}
Sampling with non-uniform probabilities was the subject of a line of research in classical compressed sensing. Seminal works involving the notion of local coherence in classical compressed sensing are~\cite{krahmerLocalCoherenceSampling, krahmerStableRobustSampling2014, puyVariableDensityCompressive2011, rauhutSparseLegendreExpansions2012}.
There have been quantities analogous to the local coherence appearing in the literature, such as Christoffel functions \cite{nevai1963geza} and leverage scores \cite{chatterjee1986influential, ma2014statistical}, which were recently introduced in machine learning in the context of, e.g., kernel-based methods \cite{erdelyi2020fourier} and deep learning \cite{adcock2022cas4dl}.

While writing this manuscript, we became aware of the recent paper~\cite{adcockCS4MLGeneralFramework2023b}. This work presents a framework for optimizing sampling in general scenarios based on the so-called \emph{generalized Christoffel function}, a quantity that admits the (squared) local coherence considered here as a particular case. Furthermore, the method we use to numerically approximate the coherence in Section 3 can be seen as a special case of that proposed in \cite{adcockCS4MLGeneralFramework2023b}. However, the results in \cite{adcockCS4MLGeneralFramework2023b} assume that $\mathcal{V}-\mathcal{V}$ is a union of low-dimensional subspaces. Hence, they are not directly applicable to the case of generative compressed sensing with ReLU networks, for which $\mathcal{V}-\mathcal{V}$ is in general only \emph{contained} in a union of low-dimensional subspaces. On the other hand, our theory explicitly covers the case of generative compressed sensing, and illustrates how sufficient conditions on $m$ leading to successful signal recovery depend on the generative network's parameters $(k,d,n)$. Second, we provide  recovery guarantees that hold with high probability, as opposed to expectation.

The present work directly improves on results from~\cite{berkCoherenceParameterCharacterizing2022} by improving the sample complexity from $n\lVert \boldsymbol{\alpha}\rVert_{\infty}^2$ to $\lVert \boldsymbol{\alpha}\rVert_2^2$ when the sampling probabilities are adapted to the generative model used. This is a sizable improvement in performance guarantees for a significant class of realistic generative models. It can be understood as extending the theory of generative compressed sensing with Fourier measurements to many realistic settings where we provide nearly order-optimal bounds on the sampling complexity. Indeed, in the context of the main result from~
\cite{berkCoherenceParameterCharacterizing2022}, ``favourable coherence'' corresponds to
$\lVert \boldsymbol{\alpha}\rVert_\infty \leq C\sqrt{ kd/n }$ where $C$ is an absolute constant. Despite the fact that the prior generated by a neural network
with Gaussian weights will have such a coherence~
\cite{berkCoherenceParameterCharacterizing2022}, we observe empirically in~\autoref{fig:main}d)
that for a trained generative model, a small number of Fourier coefficients have values close to one. In such cases, the main result from~\cite{berkCoherenceParameterCharacterizing2022} becomes vacuous while \autoref{recovery_with_generative_prior_from_unevenly_subsampled_incoherent_orthonormal_measurements} remains meaningful.

\paragraph{Notation}
\label{sec:notation}
  For any map $f$, we denote $\range(f)$ to be the range of $f$. We define the simplex $\Delta^{n-1} := \{ \boldsymbol{p} \in \reals^{n} : p_{i} \geq 0, \sum_{i=1}^{n} p_{i} = 1 \}$, and the sphere $\sphere{n} := \{ \boldsymbol{x} \in \reals^{n} : \| \boldsymbol{x} \|_{2} = 1 \}$. For any matrix $A \in \measfield^{m \times n}$, we denote its pseudo-inverse by $A^{+} \in \measfield^{n \times m}$.
 For a set $\mathcal{V} \in \field^n$ we denote its \emph{self-difference} $\mathcal{V}-\mathcal{V}:= \{\boldsymbol{v}_1-\boldsymbol{v}_2 | \boldsymbol{v}_1, \boldsymbol{v}_2 \in \mathcal{V}\}$. We define $\proj_{\mathcal{T}}: \field^n \to \field^n$ to be the orthogonal projection on to a set $\mathcal{T} \subseteq \field^n$ in the sense that $\proj_\mathcal{T}\boldsymbol{x}$ is a single element from the set $\argmin_{\boldsymbol{t} \in \mathcal{T}} \|\boldsymbol{x}-\boldsymbol{t}\|_2$. We let $[l]=\{1,\ldots,l\}$. We denote $\langle \cdot , \cdot \rangle$ to be the the canonical inner product in $\reals$ or $\complex$ depending on the context.
\section{Main Result}
\label{main_results}
We now state the main result of this paper, where we give an upper bound on the sample complexity required for signal recovery. The accuracy of the recovery is dependent on the measurement noise, the modelling error (how far the signal is from the prior), and imperfect optimization.   These are denoted, respectively, by $\boldsymbol{\eta}, \boldsymbol{x}^\perp$ and $\hat{\varepsilon}$ below.
\begin{theorem}
	\label{recovery_with_generative_prior_from_unevenly_subsampled_incoherent_orthonormal_measurements}
	Fix a $(k,d,n)$-generative network $G$, the cone $\mathcal{T} := \range(G)- \range(G)\subseteq \field^n$, the unitary matrix $F \in \measfield^{n \times n}$. Let $\boldsymbol{\alpha}$ be the vector of local coherences of $F$ with respect to $\mathcal{T}$. Let $\boldsymbol{p} := (\alpha_j^2/\lVert  \boldsymbol{\alpha}\rVert_2^2)_{j \in [n]} \in \Delta^{n-1}$ with $S \in  \mathbb{R}^{m \times  n}$ the corresponding random sampling matrix. Let $D \in  \mathbb{R}^{n \times n}$ be the diagonal matrix with entries $D_{i,i}= 1/\sqrt{p_i}$. Let $\tilde{D} := SDS^+ \in \mathbb{R}^{m \times  m}$. Let $\varepsilon > 0$. If 
 $$m \geq C \|\boldsymbol{\alpha}\|_2^2 \left( kd \log\left( \frac{n}{k} \right)+ \log\left(\frac{2}{\varepsilon}\right)\right)$$
 for $C$ an absolute constant, then with probability at least $1-\varepsilon$ over the realization of $S$ the following statement holds.

	\begin{state}
		\label{recovery_statement}
		For any choice of $\boldsymbol{x}_{0} \in \field^{n}$ and $\boldsymbol{\eta} \in \measfield^n$, let $\boldsymbol{b} := \frac{1}{\sqrt{m}}SF\boldsymbol{x} + \boldsymbol{\eta}$ and $\boldsymbol{x}^\perp := \boldsymbol{x}_{0} - \proj_{\mathcal{T}} \boldsymbol{x}_0$, and any $\hat{\boldsymbol{x}} \in \field^n, \hat{\varepsilon}>0$ satisfying $\|\frac{1}{\sqrt{m}}SDF\hat{\boldsymbol{x}} - \tilde{D}\boldsymbol{b}\|_2 \leq \min_{\boldsymbol{x} \in \mathcal{T}} \|\frac{1}{\sqrt{m}}SDF\boldsymbol{x} - \tilde{D}\boldsymbol{b}\|_{2} + \hat{\varepsilon}$. We have that
		\begin{align*}
			\|\hat{\boldsymbol{x}} - \boldsymbol{x}_0\|_2 %
			\leq \|\boldsymbol{x}^\perp\|_2 + \frac{3}{\sqrt{m}}\|SDF\boldsymbol{x}^\perp\|_2 + 3 \|\tilde{D}\boldsymbol{\eta}\|_2 + \frac{3}{2}\hat\varepsilon.
		\end{align*}
	\end{state}
\end{theorem}
\begin{remark}
	\label{rmk:characterization_of_subsampled_preconditioner}
	The matrix $\tilde{D}$ has a vector of diagonal entries $\text{Diag}(\tilde{D})$ satisfying $\text{Diag}(\tilde{D}) = S \text{Diag}(D)$.
\end{remark}
\begin{remark}
	\label{rmk:find_generalized_version}
	We give a generalization of this result to arbitrary sampling probability vectors in \autoref{generative_non-uniform_recovery}.
\end{remark}

\section{Numerics}
\label{numerics}
\begin{figure}[h]
	\centering
    \includegraphics[width=\textwidth]{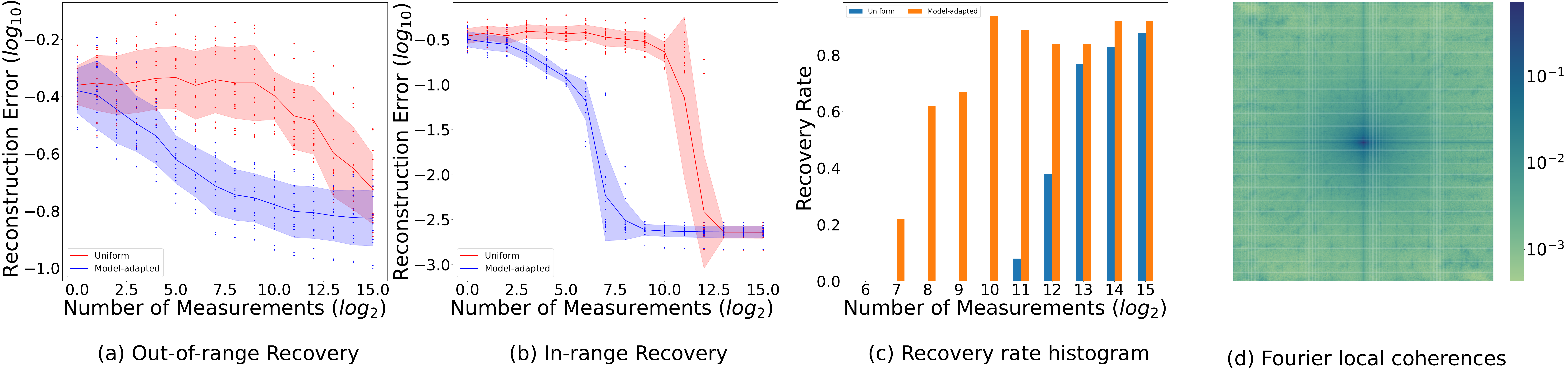}
    \caption{
    In (a) we shows the reconstruction error in terms of number of measurements with out-of-range signals.
    In (b) we present the same quantity for in-range signals.
    In (c) we treat signal recoveries with relative error less than $3\times10^{-3}$ as ``successful", and display the proportion of successful recoveries out of 64 attempts on in-range signals.
    In (d) we plot the local coherences for one channel of one image. Observe that the local coherences take on values in many different orders of magnitude and peak sharply for a small number of local coherences.
    }
    \label{fig:main}
    \vspace{-12pt}
\end{figure}

In this section, we provide empirical evidence of the connection between coherence-based non-uniform sampling and recovery error. By presenting visual (\autoref{fig:intro_comp}) and quantitative (\autoref{fig:main}) evidence, we validate that model-adapted sampling using a coherence-informed probability vector can outperform a uniform sampling scheme — requiring fewer measurements for successful recovery.
\paragraph{Coherence heuristic}
Ideally, we would compute the local coherence $\boldsymbol{\alpha}$ using ~\autoref{local_coherence}, but to our knowledge computing local coherence is intractable  for generative models relevant to practical settings \cite{berkCoherenceParameterCharacterizing2022}. Thus, we approximate the quantity by sampling points from the range of the generative model and computing the local coherence from the sampled points instead. Specifically, we sample \textit{codes} from the latent space of the generative model to generate a batch of images with shape $(B,C,H,W)$, where $B,C,H,W$ stand for batch size, number of channels, image height and image width respectively. Then, we compute the set self-difference of the image batch, and normalize each difference vector. This gives a tensor of shape $(B^2,C,H,W)$. We perform a channel-wise two-dimensional Discrete Fourier Transform (DFT) on the tensor, take the element-wise modulus, and then maximize over the batch dimension. This results in a coherence tensor $a_0$ with shape $(C,H,W).$ 
To obtain the coherence-informed probability vector of each channel, we first square element-wise, then we normalize channel-wise. 
To estimate the local coherences of our generative model we use a batch size of 5000 and 
employ the DFT from PyTorch~\cite{PyTorch}. 
\paragraph{In-Range vs Out-of-Range Signals}
We run signal recovery experiments for two kinds of images: when the signals are conditioned to be in the range of the generative model (in-range signals), and when they are directly picked from the validation set (out-of-range signals). The in-range signals are randomly generated (with Gaussian codes) by the same generative network that we use for recovery. This ensures that these signals lie within the prior set. In the out-of-range setting, a residual recovery error can be observed even with large numbers of measurements~\autoref{fig:main}. This error occurs because of the so-called model mismatch; there is some distance between the prior set and the signals.

\paragraph{Procedure for Signal Recovery}
The way we perform signal recovery goes as follows. For a given image $\boldsymbol{x}_0 \in \mathbb{R}^{C \times H \times W}$, we create a mask $M \in {\{0,1\}}^{C \times H \times W}$ 
by randomly sampling with replacement $m$ times for each channel according to the probability vector.
Let $F$ be the channel-wise DFT operator and $G:\mathbb{R}^k \to \mathbb{R}^{C \times H \times W}$ be the generative neural network (where we omit the batch dimension for simplicity). We denote $\odot$ to be the element-wise tensor multiplication and $\|\cdot\|_F$ to be the Frobenius norm.
We approximately solve the optimization program $\hat{\boldsymbol{z}} \in \argmin_{\boldsymbol{z} \in \mathbb{R}^k}\|M\odot F\boldsymbol{x}_0-M\odot FG(\boldsymbol{z})\|_F^2$ by running AdamW \cite{loshchilov2019decoupled} with $lr=0.003,\beta_1=0.9$ and $\beta_2=0.999$ for 20000 iterations on four different random initializations, and pick the code that achieves the lowest loss. The recovered signal is then $\hat{\boldsymbol{x}}:=G(\hat{\boldsymbol{z}})$. 
We measure the quality of the signal recovery by using the relative 
recovery error (rre), $\operatorname{rre}(\boldsymbol{x}_0,\hat{\boldsymbol{x}}) = \frac{\|\boldsymbol{x}_0 - \hat{\boldsymbol{x}}
\|_2}{\|\boldsymbol{x}_0\|_2}.$

Observe that \autoref{fig:main}b) demonstrates the efficiency of model-adapted sampling. Signal recovery with adapted sampling occurs with $16$ times fewer measurements than when using uniform sampling. Similar performance gains can be observed visually in \autoref{fig:intro_comp} and \autoref{fig:in-range_visual}. Comparing the number of measurements to the ambient dimension, we see from \autoref{fig:main}d) that signal recovery occurs with $\frac{m}{n} \approx 2^7/256^2=0.2\%$.

There are a few ways these numerical experiments do not directly match our theory. The sampling is done channel wise, which is technically block sampling~\cite{adcockCompressiveImagingStructure2021}. Also, the signal recovery is performed without the preconditioning factor $\tilde{D}$ that appears in~\autoref{recovery_with_generative_prior_from_unevenly_subsampled_incoherent_orthonormal_measurements}.

\section{Conclusion}
\label{conclusion}
In this paper we bring together the ideas used to quantify the compatibility of generative models with subsampled unitary measurements, which were first explored in~\cite{berkCoherenceParameterCharacterizing2022}, with ideas of non-uniform sampling from classical compressed sensing. We present the first theoretical result applying coherence-based sampling (similar to leverage score sampling, or Christoffel function sampling) to the setting where the prior is a ReLU generative neural network. We find that adapting the sampling scheme to the geometry of the problem yields substantially improved sampling complexities for many realistic generative networks, and that this improvement is significant in empirical experiments.

Possible avenues for future research include extending the theory presented in~\cite{adcockCS4MLGeneralFramework2023b} to ReLU nets by using methods introduced in the present work. This would yield the benefit of extending the theory from this paper to a number of realistic sampling schemes. A second research direction consists of investigating the optimality of the sample complexity bound that we present in this paper. The sample complexity that we guarantee includes a factor of $k^2d^2$ when the generative model is well-behaved. Whether this dependence can be reduced to $kd$, as is the case when the measurement matrix is Gaussian, is an interesting problem that remains open. Finally, the class of neural networks considered in this work could be expanded to include more realistic ones.

\newpage
\printbibliography
\newpage
\appendix
\makeatletter{}
\renewcommand\thesection{A\@arabic\c@section}
\makeatother{}
This supplementary material contains acknowledgements, a generalization of  \autoref{recovery_with_generative_prior_from_unevenly_subsampled_incoherent_orthonormal_measurements}, the proof of said generalization, properties of the piecewise linear expansion, and additional image recoveries both in-range and out-of-range.

\section{Acknowledgements}
The authors acknowledge Ben Adcock for providing feedback on a preliminary version of this paper. S. Brugiapaglia acknowledges the support of the Natural Sciences and Engineering Research Council of Canada (NSERC) through grant RGPIN-2020- 06766 and the Fonds de Recherche du Québec Nature et Technologies (FRQNT) through grant 313276.
Y. Plan is partially supported by an NSERC Discovery Grant (GR009284), an NSERC Discovery Accelerator Supplement (GR007657), and a Tier II Canada Research Chair in Data Science (GR009243). O. Yilmaz was supported by an NSERC Discovery Grant (22R82411) O. Yilmaz also acknowledges support by the Pacific Institute for the Mathematical Sciences (PIMS). 

\section{Additional Notation}
\label{sec:more_notation}
In this work we make use of absolute constants which we always label $C$. Any constant labelled $C$ is implicitly understood to be absolute and has a value that may differ from one appearance to the next. We write $a \lesssim b$ to mean $ a\leq Cb$.
We denote by $\|\cdot\|$ the operator norm, and by $\|\cdot\|_2$ the euclidean norm. 
We use capital letters for matrices and boldface lowercase letters for vectors. 
For some matrix $A \in \measfield^{m \times n}$, we denote by $\boldsymbol{a}_i^* \in \measfield^{1 \times n}$ (lowercase of the letter symbolizing the matrix) the $i^{th}$ row  vector of $A$, meaning that $A = \sum_{i=1}^m \boldsymbol{e}_i \boldsymbol{a}_i^*$. For a vector $\boldsymbol{u} \in \field^n$, we denote by $u_i$ its $i^{\text{th}}$ entry. We let $\{\boldsymbol{e}_i\}_{i \in [n]}$ be the canonical basis of $\field^n$.
\section{Generalized Main Result}
\label{generalized_main_result}
We present a generalization of \autoref{recovery_with_generative_prior_from_unevenly_subsampled_incoherent_orthonormal_measurements} which provides recovery guarantees for arbitrary sampling probabilities. To quantify the quality of the interaction between the generative model $G$, the unitary matrix $F$, and the sampling probability vector $\boldsymbol{p}$, we introduce the following quantity.

\begin{definition}
	\label{coherence_of_a_subsampled_isometry}
	Let $\mathcal{T} \subseteq \field^n$ be a cone. Let $F \in \measfield^{n \times n}$ be a unitary matrix and $\boldsymbol{p} \in \Delta^{n-1}$. Let $\boldsymbol{\alpha} \in \mathbb{R}^n$ the local coherences of $F$ with respect to $\mathcal{T}$. Then define the quanity
	$$\mu_\mathcal{T}(F, \boldsymbol{p}) := \max_{j \in [n]} \frac{\alpha_j}{\sqrt{ p_{j} }}.$$
\end{definition}

We can now state the following.

\begin{theorem}[Generalized Main Result]
	\label{generative_non-uniform_recovery}
	Fix the $(k,d,n)$-generative network $G$, the cone $\mathcal{T} := \range(G)- \range(G)\subseteq \field^n$, the unitary matrix $F \in \measfield^{n \times n}$, the probability vector $\boldsymbol{p} \in \Delta^{n-1}$, and the corresponding random sampling matrix $S \in  \mathbb{R}^{m \times  n}$. Let $D \in  \mathbb{R}^{n \times n}$ be a diagonal matrix with entries $D_{i,i}= \frac{1}{\sqrt{p_i}}$. Let $\tilde{D} := SDS^+ \in \mathbb{R}^{m \times  m}$. Let $\boldsymbol{\alpha}$ be the vector of local coherences of $F$ with respect to $\mathcal{T}$. Let $\varepsilon > 0$.

	Suppose that
	$$m \geq C \mu^2_\mathcal{T}(F, \boldsymbol{p}) \left( kd \log\left( \frac{n}{k} \right)+ \log \frac{2}{\varepsilon}\right).$$
	Furthermore, if we pick the sampling probability vector
	$$\boldsymbol{p}^* := \left(\frac{\alpha_j^2}{\lVert \boldsymbol{\alpha}\rVert_2^2}\right)_{j \in [n]},$$
	we only require that
	$$m \geq C \|\boldsymbol{\alpha}\|_2^2 \left( kd \log\left( \frac{n}{k} \right)+ \log \frac{2}{\varepsilon}\right).$$
  Then with probability at least $1-\varepsilon$ over the realization of $S$, \autoref{recovery_statement} holds.
\end{theorem}
This theorem is strictly more general than \autoref{recovery_with_generative_prior_from_unevenly_subsampled_incoherent_orthonormal_measurements}. It is therefore sufficient to prove the generalized version, which we do in the next section.

\begin{remark}
	\label{uniform_sampling_as_a_special_case_of_uneven_sampling}
	The result~\cite[Theorem 2.1]{berkCoherenceParameterCharacterizing2022} is a corollary of~\autoref{recovery_with_generative_prior_from_unevenly_subsampled_incoherent_orthonormal_measurements}; it follows from taking $\boldsymbol{p}$ as the uniform probability vector.
\end{remark}
\section{Proof of \autoref{generative_non-uniform_recovery}}

Let us first introduce the so-called Restricted Isometry Property (RIP)~\cite{candesRobustUncertaintyPrinciples2006,foucartMathematicalIntroductionCompressive2013,vershyninHighDimensionalProbabilityIntroduction2018, }.

\begin{definition}[Restricted Isometry Property]
	\label{rip}
	Let $\mathcal{T} \subseteq \field^n$ be a cone and $A \in \measfield^{m \times n}$ a matrix. We say that $A$ satisfies the Restricted Isometry Property (RIP) when
	$$\sup_{\boldsymbol{u} \in \mathcal{T}\cap \sphere{n}}\lvert \lVert A\boldsymbol{u}\rVert_2 - 1\rvert \leq \frac{1}{3}.$$
\end{definition}

Note that the constant $1/3$ is a specific choice made in order to simplify the presentation of this proof. It could be replaced by any generic absolute constant in $(0,1)$.

The following lemma says that if, conditioning on $S$, $\frac{1}{\sqrt{m}}SDF$ has the RIP on $\range(G)-\range(G)$, then we have signal recovery.
\begin{lemma}[RIP of a Subsampled and Preconditioned Matrix Yields Recovery]
	\label{rip_of_a_preconditioned_subsampled_matrix_yields_recovery}
	Let $\mathcal{V} \subseteq \field^n$ be a cone, $F \in \measfield^{n \times n}$ be a unitary matrix, $S \in  \mathbb{R}^{m \times  n}$ a matrix with all rows in the canonical basis of $\mathbb{R}^n$. Let $D \in  \mathbb{R}^{n \times n}$ be a diagonal matrix. Let $\tilde{D} := SDS^+ \in \mathbb{R}^{m \times  m}$.

	If $\frac{1}{\sqrt{m}} SDF$ has the RIP on the cone $\mathcal{T}:= \mathcal{V}-\mathcal{V}$, then \autoref{recovery_statement} holds.

\end{lemma}

See \hyperlink{proof:rip_of_a_preconditioned_subsampled_matrix_yields_recovery}{the proof} in Appendix~\autoref{proof_of_the_lemmas}.

We now proceed to prove a slightly stronger statement than what is required by \autoref{rip_of_a_preconditioned_subsampled_matrix_yields_recovery}; that the RIP holds on the piecewise linear expansion $\Delta(\range(G)-\range(G)) \supseteq \range(G)-\range(G)$.

To control the complexity of $G$, we count the number of affine pieces that it comprises. We do this with the result~\cite[Lemma A.6]{berkCoherenceParameterCharacterizing2022}, which we re-write below for convenience.

\begin{lemma}[Containing the Range of a ReLU Network in a Union of Subspaces]
	\label{lemma:counting-affine-pieces}
	Let $G$ be a $(k,d,n)$-generative network with layer widths
        $k=k_0\leq k_1, \ldots, k_{d-1} \le k_d$ where $k_d = n$ and $\bar k := \left( \prod_{\ell = 1}^{d-1} k_{\ell}\right)^{1/(d-1)}$. Then $\range(G)$ is a union of
	no more than $N$ at-most $k$-dimensional polyhedral cones where
	\begin{align*}
		\log{N} & \leq k(d-1)\log{\left( \frac{2e \bar k}{k}  \right)} \lesssim kd\log{\left(\frac{n}{k}\right)}.
	\end{align*}
\end{lemma}

From this result, we see that $\mathcal{T}:= \range(G)-\range(G)$ is contained in a union of no more than $N^2$ affine pieces each of dimension no more than $2k$. Then from~\autoref{rmk:Delta-properties} (for the proof, see~\cite[Remark A.2]{berkCoherenceParameterCharacterizing2022}), the cone $\Delta(\mathcal{T})$ is a union of no more than $N^2$ subspaces each of dimension at-most $2k$ (the factor of two will be absorbed into the absolute constant of the statement.)
Fix $\mathcal{U} \subseteq \mathcal{T}$ to be any one of these subspaces. Then the following lemma implies that the matrix $\frac{1}{\sqrt{m}}SDF$ has the RIP on $\mathcal{U}$ with high probability.

\begin{lemma}[Deviation of Subsampled Preconditioned Unitary Matrix on a Subspace]
	\label{deviation_of_unevenly_subsampled_basis_measurements_with_replacement_on_incoherent_subspace_from_matrix_bernstein}
	Let $F \in \measfield^{n \times n}$ be a unitary matrix, and $S \in  \mathbb{R}^{m \times  n}$ a random sampling matrix associated with the probability vector $\boldsymbol{p} \in  \Delta^{n-1}$. Let $D \in  \mathbb{R}^n$ be a diagonal pre-conditioning matrix with entries $D_{i,i}= \frac{1}{\sqrt{p_i}}$. Let $t > 0$. Let $\mathcal{U} \subseteq \field^n$ be a subspace of dimension $k$. Then
	\begin{equation}
		\label{eq:mat_subspace_dev}
		\sup_{\boldsymbol{x} \in \mathcal{U}\cap \sphere{n}} \left|\frac{1}{\sqrt{m}}\|SDF\boldsymbol{x}\|_{2}-1 \right|  \lesssim \frac{\mu_\mathcal{U}(F, \boldsymbol{p})}{\sqrt{ m}}\sqrt{\log k} + \frac{\mu_\mathcal{U}(F, \boldsymbol{p})}{ \sqrt{m}} t
	\end{equation}
	with probability at least $1-2\exp(-t^2)$.
\end{lemma}

See \hyperlink{proof:deviation_of_unevenly_subsampled_basis_measurements_with_replacement on incoherent subspace from matrix bernstein}{the proof} in~\autoref{proof:deviation_of_unevenly_subsampled_basis_measurements_with_replacement_on_incoherent_subspace_from_matrix_bernstein}.

Since $\mathcal{U} \subseteq \Delta(\mathcal{T})$ we have that $\mu_\mathcal{U}(F, \boldsymbol{p}) \le \mu_\mathcal{T}(F, \boldsymbol{p})$. Using this fact to upper-bound the r.h.s. of~\autoref{eq:mat_subspace_dev}, we find an identical concentration inequality that applies to each of the subspaces constituting $\Delta(\mathcal{T})$.  By using~\autoref{lemma:counting-affine-pieces} to bound the number of subspaces, we control the deviation of $\frac{1}{\sqrt{m}}SDF$ uniformly over all the subspaces constituting $\Delta(\mathcal{T})$ with a union bound. We find that, with probability at least $1-2\exp(-t^2)$,
$$\sup_{\boldsymbol{x} \in \Delta(\mathcal{T})\cap \sphere{n}} \left| \frac{1}{\sqrt{m}}\left\|  SDF\boldsymbol{x}\right\|_{2} - 1 \right|$$
$$\lesssim \frac{\mu_\mathcal{T}(F, \boldsymbol{p})}{\sqrt{ m}}\sqrt{\log k} + \frac{\mu_\mathcal{T}(F, \boldsymbol{p}) }{ \sqrt{m}} \sqrt{ kd \log\left( \frac{n}{k} \right) } + \frac{\mu_\mathcal{T}(F, \boldsymbol{p}) }{ \sqrt{m}} t.$$
For the method by which we applied the union bound, see~\cite[Lemma A.2]{berkCoherenceParameterCharacterizing2022}. In the r.h.s. of the equation above, the second term dominates the first, so the expression simplifies to
\begin{equation}
	\label{eq:cone_deviation}
	\sup_{\boldsymbol{x} \in \Delta(\mathcal{T})\cap \sphere{n}} \left| \frac{1}{\sqrt{m}}\|SDF\boldsymbol{x}\|_{2} - 1 \right| \lesssim \frac{\mu_\mathcal{T}(F, \boldsymbol{p}) }{ \sqrt{m}} \sqrt{ kd \log\left( \frac{n}{k} \right) } + \frac{\mu_\mathcal{T}(F, \boldsymbol{p}) }{ \sqrt{m}} t.
\end{equation}
By fixing $t = \sqrt{ \log\left( \frac{2}{\varepsilon} \right) }$ we find that the RIP holds with probability at least $1-\varepsilon$ on $\mathcal{T}$ when
\begin{equation}
\label{eq:first_condition_m}
m \geq C \mu_\mathcal{T}(F, \boldsymbol{p})^2 \left( kd \log\left( \frac{n}{k} \right)+ \log \frac{2}{\varepsilon}\right).
\end{equation}
Then we find that the first part of~\autoref{generative_non-uniform_recovery} follows from~\autoref{rip_of_a_preconditioned_subsampled_matrix_yields_recovery}.

The second sufficient condition on $m$ follows from picking $\boldsymbol{p}$ so as to minimize the factor $\mu_\mathcal{T}(F, \boldsymbol{p})^2$ in~\autoref{eq:first_condition_m}.

\begin{lemma}[Adapting the Sampling Scheme to the Model]
	\label{adapt_with_replacement_sampling_scheme_given_fixed_local_coherences}
	Let $F \in \measfield^{n \times n}$ be a unitary matrix, and $S \in  \mathbb{R}^{m \times  n}$ a random sampling matrix associated with the probability vector $\boldsymbol{p} \in  \Delta^{n-1}$. Let $\boldsymbol{\alpha}$ be the local coherences of $F$ with respect to a cone $\mathcal{T} \in \field^n$.
	$$\boldsymbol{p}_j^* := \left(\frac{\alpha_j^2}{\lVert  \boldsymbol{\alpha}\rVert_2^2}\right)_{j \in  [n]} \in \argmin_{\boldsymbol{p} \in \Delta^{n-1}}(\mu_\mathcal{T}(F, \boldsymbol{p})).$$
	It achieves a value of
	$$\mu_\mathcal{T}(F, \boldsymbol{p}^*) = \lVert  \boldsymbol{\alpha} \rVert_2.$$
\end{lemma}

See \hyperlink{proof:adapt_with_replacement_sampling_scheme_given_fixed_local_coherences}{the proof}.

Applying \autoref{adapt_with_replacement_sampling_scheme_given_fixed_local_coherences} to \autoref{eq:first_condition_m} concludes the proof of \autoref{generative_non-uniform_recovery}.

\section{Proof of the Lemmas}
\label{proof_of_the_lemmas}

\begin{proof}[\hypertarget{proof:rip_of_a_preconditioned_subsampled_matrix_yields_recovery}{Proof of \autoref{rip_of_a_preconditioned_subsampled_matrix_yields_recovery}}]
	\label{proof:rip_of_a_preconditioned_subsampled_matrix_yields_recovery}
	Let $F \in \measfield^{n \times n}$ be a unitary matrix, and $S \in  \mathbb{R}^{m \times  n}$ a random sampling matrix associated with the probability vector $\boldsymbol{p} \in  \Delta^{n-1}$. Let $D \in  \mathbb{R}^{n \times n}$ be a diagonal matrix with entries $D_{i,i}= \frac{1}{\sqrt{p_i}}$. Let $\tilde{D} = SDS^+$.

	We let $\tilde{\boldsymbol{b}} := \tilde{D}\boldsymbol{b}$ and $\tilde{\boldsymbol{\eta}} := \tilde{D}\boldsymbol{\eta}$. By left-multiplying the equation $\boldsymbol{b} = \frac{1}{\sqrt{m}}SF\boldsymbol{x}_0 + \boldsymbol{\eta}$ by $\tilde{D}$ we get $\tilde{\boldsymbol{b}} = \frac{1}{\sqrt{m}}SDF\boldsymbol{x}_0 + \tilde{\boldsymbol{\eta}}$. Notice that the linear operator $\frac{1}{\sqrt{m}}SDF$ has the RIP by assumption.

	By triangle inequality and the observation that $\proj_{\mathcal{T}}\boldsymbol{x}_0 \in  \mathcal{T}$,
	\begin{align*}
		\left\|\frac{1}{\sqrt{m}}SDF\hat{\boldsymbol{x}} - \tilde{\boldsymbol{b}}\right\|_2 %
		 & \leq \min_{\boldsymbol{x}\in \mathcal{T}} \left\|\frac{1}{\sqrt{m}}SDF\boldsymbol{x} - \tilde{\boldsymbol{b}}\right\|_2 + \hat\varepsilon %
		\\
		 & \leq \left\|\frac{1}{\sqrt{m}}SDF\proj_{ \mathcal{T}}\boldsymbol{x}_0 - \tilde{\boldsymbol{b}}\right\|_2 + \hat \varepsilon
		\\
		 & = \left\|\frac{1}{\sqrt{m}}SDF\boldsymbol{x}^\perp + \tilde{\boldsymbol{\eta}}\right\|_2 + \hat \varepsilon                               %
		\\
		 & \leq \frac{1}{\sqrt{m}}\|SDF\boldsymbol{x}^\perp\|_2 + \|\tilde{\boldsymbol{\eta}}\|_2 + \hat\varepsilon.
	\end{align*}
	Since $\hat{\boldsymbol{x}}, \proj_{ \mathcal{T}}\boldsymbol{x}_0 \in  \mathcal{T}$, with the RIP property we find that
	\begin{align*}
		 \left\|\frac{1}{\sqrt{m}}SDF \hat{\boldsymbol{x}} - \tilde{\boldsymbol{b}}\right\|_2                    
		 & = \left\|\frac{1}{\sqrt{m}}SDF\left(\hat{\boldsymbol{x}} - \proj_{ \mathcal{T}}\boldsymbol{x}_0\right) - \frac{1}{\sqrt{m}}SDF\left(\boldsymbol{x}_0 - \proj_{ \mathcal{T}}\boldsymbol{x}_0\right) - \tilde{\boldsymbol{\eta}} \right\|_2 %
		\\
		 & \geq \frac{1}{\sqrt{m}}\|SDF\left(\hat{\boldsymbol{x}} - \proj_{ \mathcal{T}}\boldsymbol{x}_0\right)\|_2 - \frac{1}{\sqrt{m}}\|SDF\boldsymbol{x}^\perp\|_2 - \|\tilde{\boldsymbol{\eta}}\|_2                                              %
		\\
		 & \geq \left( 1 - \frac{1}{3} \right) \| \hat{\boldsymbol{x}} - \proj_{ \mathcal{T}}\boldsymbol{x}_0 \|_2 - \frac{1}{\sqrt{m}}\|SDF\boldsymbol{x}^\perp\|_2 - \|\tilde{\boldsymbol{\eta}} \|_2.
	\end{align*}
	Assembling the two inequalities gives
	\begin{align*}
		\left\| \hat{\boldsymbol{x}} - \proj_{ \mathcal{T}}\boldsymbol{x}_0\right\|_2 \leq \frac{1}{1 - \frac{1}{3}} \left[2 \frac{1}{\sqrt{m}}\|SDF\boldsymbol{x}^\perp\|_2 + 2\|\tilde{\boldsymbol{\eta}}\|_2 + \hat\varepsilon\right].
	\end{align*}
	Finally, we apply triangle inequality to get
	\begin{align*}
		\|\hat{\boldsymbol{x}} - \boldsymbol{x}_0\|_2 %
		 & \leq \|\boldsymbol{x}_0 - \proj_{ \mathcal{T}}\boldsymbol{x}_0\|_2 + \|\hat{\boldsymbol{x}} - \proj_{ \mathcal{T}}\boldsymbol{x}_0\|_2
		\\
		 & \leq \|\boldsymbol{x}^\perp\|_2 + \frac{3}{\sqrt{m}}\|SDF\boldsymbol{x}^\perp\|_2 + 3\|\tilde{\boldsymbol{\eta}}\|_2 + \frac{3}{2}\hat\varepsilon.
	\end{align*}
\end{proof}

\begin{proof}[\hypertarget{proof:deviation_of_unevenly_subsampled_basis_measurements_with_replacement_on_incoherent_subspace_from_matrix_bernstein}{Proof of \autoref{deviation_of_unevenly_subsampled_basis_measurements_with_replacement_on_incoherent_subspace_from_matrix_bernstein}}]
	\label{proof:deviation_of_unevenly_subsampled_basis_measurements_with_replacement_on_incoherent_subspace_from_matrix_bernstein}
  In what follows, we will use that $\forall \boldsymbol{x} \in  \mathcal{U}$, $SDF\boldsymbol{x} = SDFP_\mathcal{U}^*P_\mathcal{U}\boldsymbol{x}$ where $P_\mathcal{U} \in \field^{k \times n}$ is the matrix with rows chosen to be any fixed orthonormal basis of $\mathcal{U}$. Indeed, notice that $P_\mathcal{U}^* P_\mathcal{U} = \Pi_\mathcal{U} \in \field^{n \times n}$, the orthonormal projection on to $\mathcal{U}$. Now consider
	\begin{align*}
		(\star) :=\sup_{\boldsymbol{x} \in \mathcal{U} \cap \sphere{n}} \left| \frac{1}{m}\left\| SDF\boldsymbol{x}\right\|_{2}^2 - 1 \right|
		 & = \sup_{\boldsymbol{x} \in \mathcal{U} \cap \sphere{n}} \left| \frac{1}{m}\left\| S D F P_\mathcal{U}^* P_\mathcal{U} \boldsymbol{x}\right\|_{2}^2 - 1 \right|                        \\
     & = \sup_{\boldsymbol{u} \in \field^k \cap \sphere{k}} \left| \frac{1}{m}\left\| S D F P_\mathcal{U}^* \boldsymbol{u}\right\|_{2}^2 - 1 \right|                                  \\
     & = \frac{1}{m}\sup_{\boldsymbol{u} \in \field^k \cap \sphere{k}} \left| \boldsymbol{u}^* \left[ (S D F P_\mathcal{U}^*)^* (S D F P_\mathcal{U}^*) - mI \right]\boldsymbol{u} \right|. \\
	\end{align*}
	The second equality above follows from a change of variables $P_\mathcal{U} \boldsymbol{x} \to  \boldsymbol{u} \in \field^k$. Since the matrix within the square bracket is symmetric, the last expression we find above corresponds to an operator norm.
 \begin{align}
	(\star)&=\frac{1}{m}\left\|P_\mathcal{U} F^* D S^*  S D F P_\mathcal{U}^* - mI\right\| \\
		\label{eq:thing_to_bound}
		&= \frac{1}{m}\left\|\sum_{i=1}^{m} \left[P_\mathcal{U} F^* D \boldsymbol{s}_i \boldsymbol{s}_i^* D F P_\mathcal{U}^* - I\right]\right\|
	\end{align}
	This is a sum of independent random matrices because the sampling matrix matrix $S$ is random and has independent rows. We now consider what will be the central ingredient of this proof: the Matrix Bernstein concentration bound~\cite[Theorem 5.4.1]{vershyninHighDimensionalProbabilityIntroduction2018}. We will use it to bound \autoref{eq:thing_to_bound}.

	\begin{lemma}[Matrix Bernstein]
		\label{matrix_bernstein}
		Let $X_1, ..., X_N$ be independent, mean zero, $n \times n$ symmetric random matrices, such that $||X_i|| \leq K$ almost surely for all i. Then, for every $t\geq 0$, we have
		$$\mathbb{P} \left\{ \left\lVert \sum_{i=1}^N X_i \right\rVert \geq t \right\} \leq 2n\exp \left( - \frac{t^2/2}{\sigma^2 + Kt/3}\right),$$
		where $\sigma^2=\left\|\sum_{i=1}^N \mathbb{E}X_i^2\right\|$.
	\end{lemma}

  To compute $\sigma^2$ and $K$, we notice that we can write 
  $$\sum_{i=1}^{m} \left[P_\mathcal{U} F^* D \boldsymbol{s}_i \boldsymbol{s}_i^* D F P_\mathcal{U}^* - I\right] = \sum_{i \in [n]} [\boldsymbol{v}_i \boldsymbol{v}_i^* -I]$$
  for the random vectors $\boldsymbol{v}_i := P_\mathcal{U} F^* D \boldsymbol{s}_i$. These vectors have two key properties. First, they are \emph{isotropic}; this is the property that
	\begin{align*}
		\mathbb{E}[\boldsymbol{v}_i \boldsymbol{v}_i^*] & = \mathbb{E} [P_\mathcal{U} F^* D \boldsymbol{s}_i \boldsymbol{s}_i^* D F P_\mathcal{U}^* ]    \\
		& = P_\mathcal{U} F^* D \mathbb{E}[\boldsymbol{s}_i \boldsymbol{s}_i^*] D F P_\mathcal{U}^* = I.
		\end{align*}
	The isotropic property gives us immediately that, as required, the matrices $\{\boldsymbol{v}_i \boldsymbol{v}_i^* - I\}_{i \in [m]}$ are mean-zero.

	The second property of the vectors $\{\boldsymbol{v}_i\}_{i\in [m]}$ is that they have bounded magnitude almost surely.
	\begin{align*}
		\|\boldsymbol{v}_i\|_2 & = \|P_\mathcal{U} (F^* D \boldsymbol{s}_i)\|_2                                                                \\
		          & = \frac{1}{\sqrt{p_i}}\|P_\mathcal{U} \boldsymbol{f}_i\|_2                                                    \\
		          & = \frac{1}{\sqrt{p_i}}\sup_{\boldsymbol{x} \in \mathcal{U} \cap \sphere{n}} \left\langle \boldsymbol{x}, \boldsymbol{f}_i \right\rangle \\
		          & \leq \mu_\mathcal{U}(F, \boldsymbol{p}).
	\end{align*}
	Let $\mu := \mu_\mathcal{U}(F, \boldsymbol{p})$ for conciseness.
	We proceed to compute a value for $K$. By triangle inequality and property of the operator norm of rank one matrices, we see that
	$$\|\boldsymbol{v}_i \boldsymbol{v}_i^* - I\| \le  \|\boldsymbol{v}_i\|_2^2 + 1 \le 2\mu^2.$$

	The last inequality holds because of the lower bound $\mu^2 \ge 1$, which we now justify. Consider that from \autoref{adapt_with_replacement_sampling_scheme_given_fixed_local_coherences} we have that $\mu \ge \|\boldsymbol{\alpha}\|_2$, and furthermore that for any fixed one-dimensional subspace $\mathcal{U}_0 \subseteq \mathcal{U}$, we have that $\|\boldsymbol{\alpha}\|_2 =\|F \hat{\boldsymbol{u}}\|_2 = 1$ for a unit vector $\hat{\boldsymbol{u}} \in \mathcal{U}_0$. This gives us the desired lower bound by monotonicity of $\mu$ over set containment.

	We now compute $\sigma^2,$ similarly \cite[Lemma 12.21]{adcockCompressiveImagingStructure2021}:
	Then
	\begin{align*}
		\sigma^2 & = \left\|\sum_{i=1}^m \mathbb{E}\left[(\boldsymbol{v}_i \boldsymbol{v}_i^* - I)^2\right] \right \| \\
		         & = \sup_{\boldsymbol{u} \in \field^k \cap \sphere{k}}\left\langle \boldsymbol{u}, \sum_{i=1}^m (\mathbb{E}[\boldsymbol{v}_i \boldsymbol{v}_i^*\boldsymbol{v}_i \boldsymbol{v}_i^*]-I) \boldsymbol{u} \right\rangle \\
		         & = \sup_{\boldsymbol{u} \in \field^k \cap \sphere{k}}\left\langle \boldsymbol{u}, \sum_{i=1}^m \lVert \boldsymbol{v}_i\rVert_{2}^2 \mathbb{E}[\boldsymbol{v}_i\boldsymbol{v}_i^*] \boldsymbol{u}\right\rangle - \sum_{i=1}^m \|\boldsymbol{u}\|_2^2     \\
		         & \le \sup_{\boldsymbol{u} \in \field^k \cap \sphere{k}}\left\langle \boldsymbol{u}, \sum_{i=1}^m \mu^2 I \boldsymbol{u}\right\rangle - \sum_{i= 1}^m \lVert \mathbb{E}[\boldsymbol{v}_i \boldsymbol{v}_i^*] \boldsymbol{u} \rVert_{2}^2 \\
		         & \le  \mu^2 m.
	\end{align*}
	The second equality holds because the matrix is symmetric non-negative definite, and the last inequality is obtained by dropping the second negative term.

	Then applying the Matrix Bernstein yields

	$$\mathbb{P}\left\{ \left\|\sum_{i=1}^{m} \left[P_\mathcal{U} F^* D \boldsymbol{s}_i \boldsymbol{s}_i^* D F P_\mathcal{U}^* - I\right]\right\| \ge  t \right\}
		\leq 2k \exp\left( -\frac{t^2 /2}{\mu^2m + 2\mu^2 \frac{t}{3}} \right).$$
	Substituting with \autoref{eq:thing_to_bound}, we get
	$$\mathbb{P}\left\{ \sup_{\boldsymbol{x} \in \mathcal{U} \cap \sphere{n}} \left| \frac{1}{m}\left\| SDF\boldsymbol{x}\right\|_{2}^2 - 1 \right| \ge  \frac{t}{m} \right\}
		\leq 2k \exp\left( -\frac{t^2 /2}{\mu^2m + 2\mu^2 \frac{t}{3}} \right).$$
	We would like to get our result in terms of the $l_{2}$ norm without the square. For this purpose we make use of the ``square-root trick'' that can be found in \cite[Theorem 3.1.1]{vershyninHighDimensionalProbabilityIntroduction2018}. We re-write the above as
	$$\mathbb{P}\left\{ \sup_{\boldsymbol{x} \in \mathcal{U} \cap \sphere{n}} \left| \frac{1}{m}\left\| SDF\boldsymbol{x}\right\|_{2}^2 - 1 \right| \ge \frac{t}{m} \right\}
		\leq 2k \exp\left( - C \min\left(\frac{t^2 }{\mu^2 m}, \frac{t }{\mu^2}\right) \right).$$
	We make the substitution $t \to m\max(\delta, \delta^2)$, which yields
	$$\mathbb{P}\left \{ \sup_{\boldsymbol{x} \in \mathcal{U} \cap \sphere{n}} \left| \frac{1}{m}\|SDF \boldsymbol{x}\|_2^2 - 1 \right| \geq \max(\delta, \delta^2) \right \} \leq 2k \exp\left( -C \frac{m\delta^2 }{\mu^2} \right).$$
	With the restricted inequality $\forall  a,\delta >0,|a -1| \ge \delta \implies |a^2 -1| \ge \max(\delta, \delta^2) $, we infer that
	$$\mathbb{P}\left\{ \sup_{\boldsymbol{x} \in \mathcal{U} \cap \sphere{n}} \left| \frac{1}{\sqrt{m}}\|SDF\boldsymbol{x}\|_{2} - 1 \right|  \geq \delta \right\} \leq 2k \exp\left( -C \frac{m\delta^2}{\mu^2} \right).$$
	Finally, with another substitution $\left(\frac{cm\delta^2}{\mu^2} - \log k\right) \to t^2$ we write that
	$$\sup_{\boldsymbol{x} \in \mathcal{U}\cap \sphere{n}} \left|\frac{1}{\sqrt{m}}\|SDF\boldsymbol{x}\|_{2}-1 \right|  \lesssim \frac{\mu}{\sqrt{ m}}\sqrt{\log k} + \frac{\mu}{ \sqrt{m}} t$$
	with probability at least $1 - 2\exp(-t^2)$.
\end{proof}

\begin{proof}[\hypertarget{proof:adapt_with_replacement_sampling_scheme_given_fixed_local_coherences}{Proof of \autoref{adapt_with_replacement_sampling_scheme_given_fixed_local_coherences}}]
	\label{proof:adapt_with_replacement_sampling_scheme_given_fixed_local_coherences}
	It suffices to show that $\boldsymbol{p}^* := \left(\frac{\alpha_j^2}{\|\boldsymbol{\alpha}\|_2^2}\right)_{j \in [n]}$ satisfies
	$$\boldsymbol{p}^* \in \argmin_{\boldsymbol{p} \in \Delta^{n-1}} \mu^2_\mathcal{T}(F, \boldsymbol{p}) = \argmin_{\boldsymbol{p} \in \Delta^{n-1}} \max_{j \in [n]} \frac{\alpha_j^2}{p_j }.$$

	The vector $\boldsymbol{p}^*$ achieves a value of
	$$\mu^2_\mathcal{T}(F, \boldsymbol{p}^*) = \|\boldsymbol{\alpha}\|_2^2$$
	which is the minimum. Indeed, for any fixed vector $\boldsymbol{p} \in \Delta^{n-1}$,

	\begin{equation*}
		\label{eq:cvx_cmb}
		\max_{j \in [n]} \frac{\alpha_j^2}{p_j} \ge \sum_{j=1}^{n} a_j \frac{\alpha_j^2}{p_j} \quad \forall \boldsymbol{a} \in \Delta^{n-1}.
	\end{equation*}
	The inequality above holds because the r.h.s. is a convex combination of the terms $\left\{\frac{\alpha_j^2}{p_j}\right\}_{j \in [n]}$, and is therefore upper-bounded by the maximum element of the combination. By letting $\boldsymbol{a} = \boldsymbol{p}$, we get
	$$\max_{j \in [n]} \frac{\alpha_j^2}{p_j} \ge \|\boldsymbol{\alpha}\|_2^2.$$
	Therefore, $\boldsymbol{p}^* \in \argmin_{\boldsymbol{p} \in \Delta^{n-1}} \mu^2_\mathcal{T}(F, \boldsymbol{p})$.
\end{proof}
\section{Properties of the Piecewise Linear Expansion}
The following is a subset of the elements in remark~\cite[Remark A.2]{berkCoherenceParameterCharacterizing2022}, to which we refer the reader for the proof.
\begin{remark}[Properties of the Piecewise Linear Expansion]
  \label{rmk:Delta-properties}
  Below we list several properties about $\Delta$. Let
  $\mathcal{C} = \bigcup\limits_{i=1}^N\mathcal{C}_i$ be the union of
  $N\in\nats$ convex cones $\mathcal{C}_{i}$.
  \begin{enumerate}
  \item The set $\Delta(\mathcal{C})$ is uniquely defined. In particular, it is
    independent of the (finite) decomposition of $\mathcal{C}$ into convex
    cones.
  \item If $\max_{i \in [N]}\dim \mathcal{C}_{i} \leq k$, then
    $\Delta(\mathcal{C})$ is a union of no more than $N$ at-most $k$-dimensional
    linear subspaces.
  \item The set $\Delta(\mathcal{C})$ satisfies
    $\mathcal{C} \subseteq \Delta(\mathcal{C}) \subseteq
    \mathcal{C}-\mathcal{C}$.
  \item There are choices of $\mathcal{C}$ for which
    $\mathcal{C} \subsetneq \Delta(\mathcal{C})$ (for instance, refer to the
    example at the end of this section).
\end{enumerate}
\vspace{-12pt}
\end{remark}

\section{Experimental Specifications}
\paragraph{CelebA with RealnessGAN}
CelebFaces Attributes Dataset (CelebA) is a dataset with over 200,000 celebrity face images \cite{liu2015faceattributes}. We train a model on most images of the CelebA dataset, leaving out 2000 images to comprise a validation set. We crop the colour images to 256 by 256, leading to 256 $\times$ 256 $\times$ 3 = 196608 pixels per image. On this dataset, we train a RealnessGAN with the same training setup as described in \cite{xiangli2020real}, substituting the last Tanh layer with  \href{https://pytorch.org/docs/stable/generated/torch.nn.Hardtanh.html}{{HardTanh}}, a linearized version of Tanh, to fit in our theoretical framework. See \cite{xiangli2020real} for more training and architecture details.
\newpage
\section{Additional Image Recoveries}

\begin{figure}[h]
    \centering
\includegraphics[width=\linewidth]{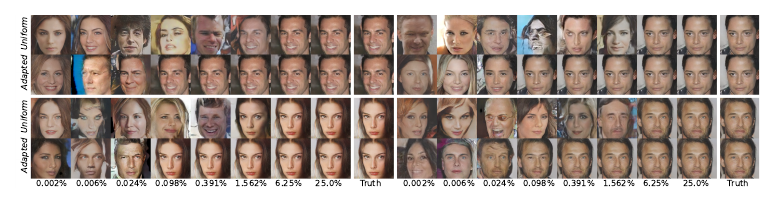}
\caption{In-range signal recovery of images using uniform sampling and model-adapted sampling. The sampling rate on the bottom is computed using the ratio between number of measurements and number of pixels. }
    \label{fig:in-range_visual}
\end{figure}

\begin{figure}[h]
    \centering
\includegraphics[width=\linewidth]{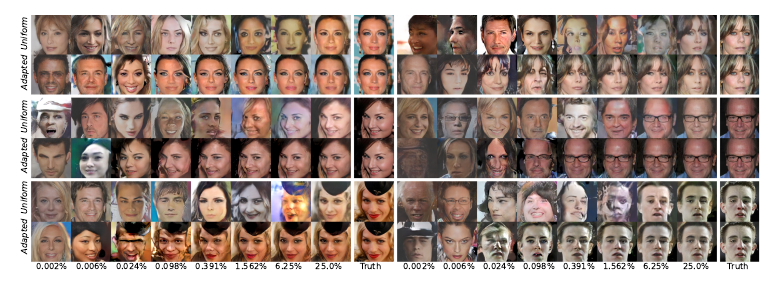}
    \caption{Out-of-range signal recovery of images using uniform sampling and model-adapted sampling. }
    \label{fig:out-range_visual}
\end{figure}

\end{document}